\documentclass[fleqn,twoside]{article}
\usepackage{espcrc2}
\usepackage{epsfig}

\title{Generation of confinement and other nonperturbative
effects by infrared gluonic degrees of freedom}

\author{Michael Engelhardt\address{Physics Department,
        New Mexico State University, \\ 
        Las Cruces, NM 88003, USA}
        }
       
\begin{document}

\begin{abstract}
Recent progress in understanding the emergence of confinement and other
nonperturbative effects in the strong interaction vacuum is reviewed.
Special emphasis is placed on the role of different types of collective
infrared gluonic degrees of freedom in this respect. After a survey of
complementary approaches, models of the QCD vacuum based on center vortices,
Abelian magnetic monopoles and topological charge lumps such as instantons,
merons and calorons are examined. Both the physical mechanisms governing
these models as well as recent lattice studies of the respective degrees of
freedom are reviewed.
\vspace{-0.1cm}
\end{abstract}

\maketitle

\section{INTRODUCTION}
Strong interaction physics is characterized by a range of nonperturbative
phenomena. Quarks and gluons are confined into color singlet hadrons.
Chiral symmetry is realized in the Goldstone mode, which decisively
influences low-energy hadronic physics through the associated (quasi-)
\linebreak
Goldstone bosons, i.e., the pions. On the other hand, the axial $U_A (1)$
part of the flavor symmetry is broken by an anomaly which
manifests itself e.g.~in the large mass of the corresponding flavor singlet
pseudoscalar $\eta^{\prime } $ meson. This review is concerned with recent
progress in understanding the QCD vacuum structure which
leads to the emergence of such phenomena. While the primary emphasis is on
confinement, also the other nonperturbative effects highlighted above are
taken into account; a cogent picture of the QCD vacuum should provide a
comprehensive explanation of these phenomena on a unified footing, rather
than a collection of separate mechanisms.

This review particularly focuses on lattice studies of the role of different 
types of collective infrared gluonic degrees of freedom in generating the
above nonperturbative effects. A variety of such degrees of freedom have
been considered, which can be roughly classified according to their
dimensionality in four-dimensional space-time:
\begin{itemize}
\item
{\bf Center vortices:}
Two-dimensional world-surfaces of quantized chromomagnetic flux.
\item
{\bf Abelian magnetic monopoles:}
One-dimensional world-lines of sources and sinks of chromomagnetic flux.
\item
{\bf Instantons, Merons, Calorons:}
Localized (i.e., zero-dimensional) lumps of topological charge
(for calorons, this classification
is not as clear-cut as for instantons and merons, since calorons generically
consist of several distributed constituents).
\end{itemize}
In keeping with the review character of this contribution, before proceeding
with the discussion of the aforementioned collective infrared degrees of
freedom, a survey of complementary recent lattice investigations of confinement
is in order. The reader's attention is also drawn to the recent review
article by Greensite \cite{greensrev}, which the present contribution has
significant overlap with, in particular as far as center vortices and
Abelian monopoles are concerned; \cite{greensrev} does not discuss
the topological charge lumps mentioned above, but on the other hand delves
more deeply into some of the subjects which are only touched upon in the
following survey.

\section{SURVEY OF RELATED WORK}
The following are the pertinent lines of investigation on which there has
been hep-lat archive posting activity since the previous Lattice Symposium
(the author has made every effort to make this survey complete in that sense):
\begin{itemize}
\item
{\bf Confinement in the Coulomb gauge:} Confining behavior of the gluon
propagator in Coulomb gauge has been shown to be linked to a remnant gauge
symmetry being realized; an order parameter associated with this symmetry
was introduced and studied \cite{goz}. The color Coulomb potential induced
by the Coulomb propagator, which is an upper bound for the static
potential between color sources, is linear both in the confined and the
high-temperature deconfined phase of $SU(2)$ Yang-Mills theory \cite{goz}.
The Coulomb propagator, the ghost form factor and the color Coulomb
potential have moreover been investigated in \cite{moyaerts}.
\item
{\bf Confinement in the Landau gauge, Kugo-Ojima confinement criterion:}
Extensive studies of the gluon propagator, the ghost propagator and the
associated running coupling have been performed in the Landau gauge
\cite{furna,bowman,taurines,gattnar}. One of these studies \cite{furna}
in particular evaluated the Kugo-Ojima confinement parameter, obtaining
the value -0.83. Another investigation \cite{bowman} focused on the
effects of dynamical quarks, finding a reduction of the infrared enhancement
of the gluon propagator. Evidence for violation
of reflection positivity in the gluon propagator, interpreted as a
manifestation of confinement, was reported in a large-volume three-dimensional
study \cite{taurines}. The Landau gauge properties in an ensemble generated
by removing center projection vortices from the full $SU(2)$ Yang-Mills
ensemble were also considered \cite{gattnar}. In this case, the
signatures of confinement in Landau gauge propagators disappear.
\item
{\bf Confinement in models with exceptional gauge groups:}
The study of confinement in models with exceptional gauge
groups is particularly interesting due to the fact that some gauge groups
exclude certain confinement mechanisms because of their internal topology.
The confining $G(2)$ theory \cite{confg2} has a trivial
center and no center vortices ($G(2)$ is its own universal covering group
and its first homotopy group is trivial, $\Pi_{1} [G(2)] = \{ 0\} $).
Moreover, by introducing an
appropriate Higgs field, the $G(2)$ gauge symmetry can be ``broken'' to
$SU(3)$ and the transition between exceptional and ordinary confinement can
be studied. Also the $Sp(2)$ and $Sp(3)$ models have been considered
\cite{confsp}. In the case of $Sp(2)$ in 2+1 dimensions, the
Svetitsky-Yaffe conjecture is applicable and confirmed numerically.
\item
{\bf SU(2) vs.~SO(3) gauge groups:} Comparing models with the $SU(2)$ and
the $SO(3)$ gauge groups is instructive, since they obey the same algebra
and only differ in their center (the $SO(3)$ gauge group, despite its trivial
center, nevertheless supports the equivalent of $SU(2)$ center vortices; the
universal covering group of $SO(3)$ is $SU(2)$, with the center $Z(2)$, and
$\Pi_{1} [SU(2)/Z(2)] = Z(2) $). The correspondence between the two models
was elucidated in detail \cite{fojaso3}, where an analytic path connecting
$SO(3)$ and $SU(2)$ lattice gauge theory at weak coupling was given. A recent
study \cite{muepreu} focused on the deconfinement transition in a modified
$SO(3)$ model, finding a correspondence of critical exponents to the $SU(2)$
theory; also Abelian magnetic monopole condensation was considered in
\cite{muepreu}, and found to be correlated with the presence of confinement.
\item
{\bf Higher representation Wilson loops:} An important characteristic of
the Yang-Mills vacuum, useful to constrain vacuum models,
is the behavior of higher representation Wilson loops. Direct evidence
has been presented that the spectrum of string tensions in $SU(3)$ Yang-Mills
theory behaves according to n-ality \cite{panag}. Also higher numbers
of colors have been investigated \cite{lucini}, finding k-string tensions
lying between the MQCD and Casimir scaling conjectures.
\item
{\bf Connecting short to long scales by decimations:} A representation of
the Yang-Mills partition function in terms of successive decimations has
been derived \cite{tombdec}. In this way, the short and long distance
regimes of $SU(N)$ lattice gauge theory can be connected, with the
aim of deriving exact statements about confinement.
\item
{\bf String breaking:} Breaking of the adjoint chromoelectric string
at large distances has been observed in 2+1-dimensional $SU(2)$ Yang-Mills
theory without resorting to a two-channel analysis \cite{kratoch},
by exploiting a noise reduction method of the L\"uscher-Weisz type
\cite{luewei} to measure Wilson loops. Also
fundamental string breaking in the presence of dynamical quarks at
finite temperatures has been studied \cite{stribre}.
\item
{\bf String-like behavior of the chromoelectric flux tube:} Sparked by
the introduction of the L\"uscher-Weisz noise reduction technique
\cite{luewei}, which permits the evaluation of
Wilson loops with unprecedented accuracy, this has been the
most active of the lines of investigation surveyed here. A number of
high precision studies of the static quark-antiquark potential have been
undertaken \cite{lwpot,juge1,juge2,caselle1,caselle2,majum1,majum2} in both
2+1 and 3+1 dimensions, using $SU(2)$, $SU(3)$, $Z(2)$ and compact $U(1)$ gauge
groups. The aim of these studies lies in extracting subleading contributions
to the potential, such as the L\"uscher term, in order to search for
string-like behavior of the chromoelectric flux tube. Bosonic string
characteristics are found. In addition, the analogue of the L\"uscher term
in the baryonic case has been computed \cite{jahnbary} in 2+1 dimensions.
\end{itemize}

\section{COLLECTIVE INFRARED GLUONIC DEGREES OF FREEDOM}
The following questions are a useful guide in considering the relevance
of the different types of infrared gluonic degrees of freedom which have
been studied:
\begin{itemize}
\item
{\bf Can the degrees of freedom in question generate the nonperturbative
effects characterizing the strong interaction in the infrared? How?}
If a particular set of degrees of freedom is to faithfully represent
the infrared structure of the QCD vacuum, it should ideally supply a
comprehensive account of the effects induced by that vacuum, rather than
explaining only particular aspects.
\item
{\bf Is it realistic to assume them to be weakly coupled/correlated?}
Infrared QCD vacuum structure can in principle be expanded in any
suitably complete set of degrees of freedom; a relevant set should satisfy
a more stringent criterion, namely that it generate the nonperturbative
phenomena characterizing the strong interaction on the basis of a weakly
coupled dynamics. Strong correlations between degrees of freedom in a given
set are a signature that they are really combined into a different set of
collective degrees of freedom which more faithfully represent the QCD vacuum.
\item
{\bf How can they be studied in lattice gauge theory and what are the
results of such studies?}
To further constrain the description of vacuum structure, it is useful
to find ways of identifying different sets of infrared gluonic degrees
of freedom in lattice gauge configurations and thus be able to investigate
their structure and relevance directly in the full theory, complementing
model studies which have to rely on their phenomenological results to
justify the use of a particular set of degrees of freedom.
\end{itemize}

\subsection{Center vortices}
In four-dimensional space-time, center vortices are represented by closed
two-dimensional world-surfaces (with a transverse thickness related to
the QCD scale). Their flux is quantized; for $SU(N)$ color, there are $N-1$
types of flux, characterized by yielding one of the $N-1$ different
nontrivial center elements of $SU(N)$ when a Wilson loop encircling
the vortex flux is evaluated. The center vortex picture of the strong
interaction vacuum assumes the vortex world-surfaces to be weakly
correlated, random surfaces on infrared length scales.

\subsubsection{Confinement}
\begin{figure}
\hspace{2.1cm} \epsfig{file=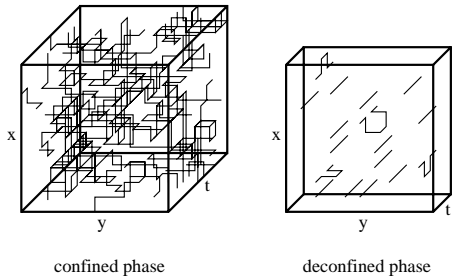,width=3cm} \vspace{-0.5cm}
\caption{Typical vortex configurations in the confined and deconfined
phases; one spatial coordinate is kept fixed.}
\vspace{-0.2cm}
\label{intui}
\end{figure}

\begin{figure}
\vspace{-0.6cm}
\hspace{4.3cm} \epsfig{file=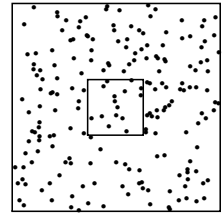,width=2.1cm}
\vspace{0.3cm}
\caption{Confinement induced by random linkings of vortices with Wilson
loops. A two-dimensional plane of area $L^2 $ containing a Wilson loop
spanning an area $A$ is pierced by vortices at random locations.}
\vspace{-0.2cm}
\label{heurist}
\end{figure}

On this basis,
both the confined and the deconfined phases can be understood; a
particularly intuitive picture is obtained by considering a slice of
the universe in which one of the spatial coordinates is kept constant,
cf.~Fig.~\ref{intui}.
At finite temperatures, when the time direction of (Euclidean) space-time
becomes too short, the vortices cannot fluctuate appreciably in the spatial
directions before closing on themselves by virtue of the periodic
boundary conditions. They extend chiefly in the time and one space
direction and cease to percolate (in the sliced universe depicted in
Fig.~\ref{intui}). As will be seen presently, percolation
is a prerequisite for confinement. The deconfining transition can be
understood as a percolation transition induced by the change in entropy
of the random vortex surfaces as the time coordinate is shortened.
To elucidate the emergence of confinement in a percolating vortex
ensemble, consider the following heuristic picture, cf.~Fig.~\ref{heurist}.
Given $N$ random vortex piercings of a plane of area $L^2 $, the probability
that $n$ of the piercings occur within an area $A$ spanned by a
Wilson loop is binomial. Considering for definiteness the $SU(2)$ case, each
piercing multiplicatively contributes a center element factor $-1$ to the
Wilson loop, i.e., for $n$ piercings within the loop, it takes the value
$(-1)^n $. Summing over all possibilities with the proper binomial weight
yields
\[
\langle W \rangle = \sum_{n=0}^{N} (-1)^{n}
\left( \begin{array}{c} N \\ n \end{array} \right)
\left( \frac{A}{L^2 } \right)^{n}
\left( 1 - \frac{A}{L^2 } \right)^{N-n}
\]
\[
\hspace{0.6cm} = \left( 1 - \frac{2\rho A}{N} \right)^{N}
\stackrel{N\rightarrow \infty }{\longrightarrow }
\exp (-2\rho A)
\]
where $L^2 $ has been eliminated in favor of the planar density
$\rho =N/L^2 $, and in the final step, the limit of a large universe
$N\rightarrow \infty $ at constant $\rho $ is taken. Thus, one obtains
an area law for the Wilson loop, with the string tension determined by
the vortex density. The crucial assumption in this argument is the
independence of the piercing points. This can only be accurate if the
vortices percolate as in the left panel of Fig.~\ref{intui}. In the
absence of percolation, there is an upper bound on the size of a vortex
cluster. As a consequence, piercing points come in pairs less than this
maximal size apart; if a vortex pierces a plane in one direction, it
must return to pierce it again in the other direction, because it must
ultimately close. Evaluating the Wilson loop in complete analogy
to above, only with piercing point {\em pairs} distributed on the plane,
gives a perimeter law. Note furthermore that adjoint Wilson loops yield
zero string tension, since they have unit value even when encircling
a single center vortex; the center vortex picture yields the correct
n-ality dependence of string tensions by construction.

\subsubsection{Topology}
\begin{figure}
\vspace{0.1cm}
\epsfig{file=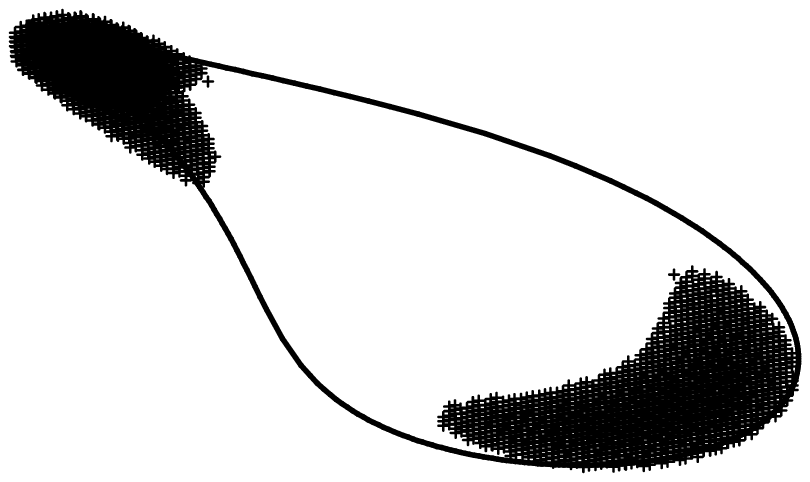,width=2cm} \hspace{0.5cm}
\epsfig{file=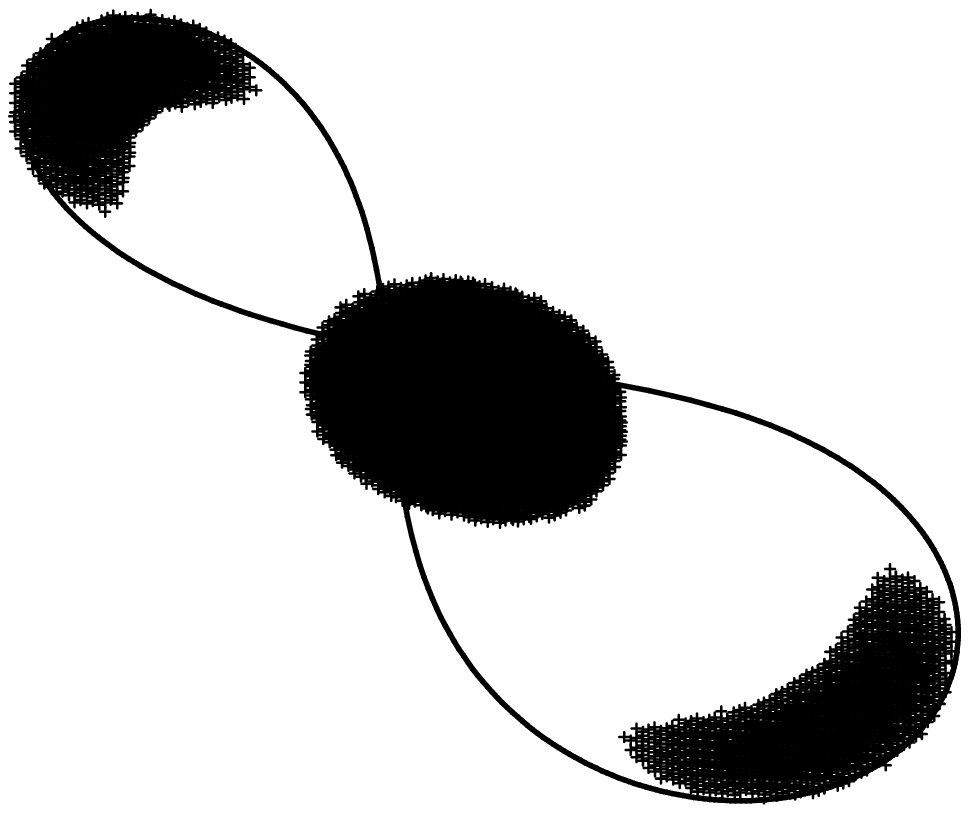,width=2cm} \hspace{0.5cm}
\epsfig{file=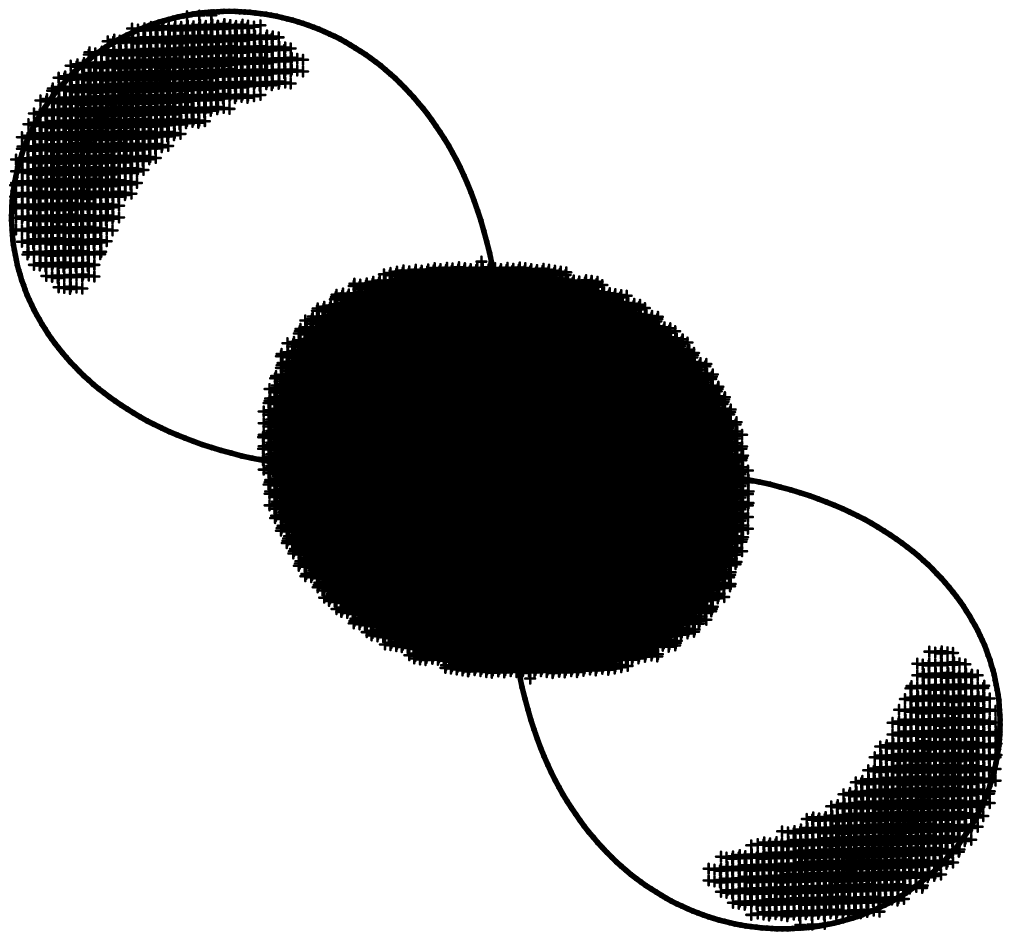,width=2cm}
\vspace{-1.2cm}
\caption{Topological charge density on a writhing, self-intersecting
vortex surface. The thin line is to guide the eye as to the location of
the geometrical center of the (thick) vortex; this is a vortex loop (left)
which is twisted until it becomes a figure eight and self-intersects (right).
The scatter plot indicates locations where the topological charge density
exceeds a certain threshold. In the center and right-hand figures, the
signature of the self-intersection is evident; on the other hand, the
kidney-shaped contributions at the periphery originate from vortex writhe
\cite{bren}.}
\vspace{-0.3cm}
\label{topex}
\end{figure}

Topological charge density is generated by vortex self-intersections and
vortex writhe. In the $SU(2)$ case, self-intersections carry topological
charge $\pm 1/2$, whereas topological charge density due to writhe
in general is distributed continuously along vortex surfaces. An
instructive example illustrating both types of contribution was presented
by F.~Bruckmann and the author in \cite{bren}, cf.~Fig.~\ref{topex}.
In the case of surfaces made up of elementary squares on a hypercubic
lattice, contributions of writhe become discrete, concentrated on lattice
sites, with modulus less than 1/2 at each site. In realistic vortex ensembles,
cf.~section \ref{randmod}, the contributions of vortex writhe to the
topological charge are statistically far more important than the ones
due to vortex self-intersections.

\begin{figure}
\vspace{0.2cm}
\hspace{0.8cm}
\epsfig{file=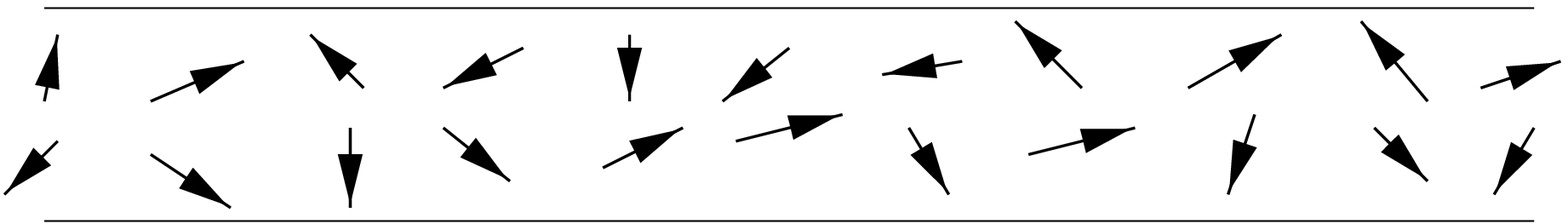,width=6cm}
\vspace{0.35cm}

\hspace{0.8cm}
\epsfig{file=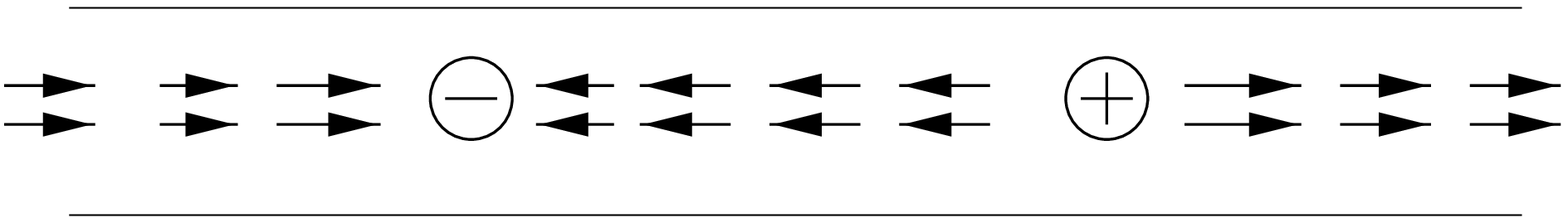,width=6cm}
\vspace{-0.65cm}
\caption{Magnetic monopoles distributed along vortices in Abelian gauges
due to vortex nonorientability (from \cite{ambmon}).}
\vspace{-0.2cm}
\label{vormon}
\end{figure}

Globally, generic vortex world-surfaces are nonorientable. If one attempts
to rotate the associated field strength into an Abelian gauge, then,
as one orients the field, say, into the positive 3-direction in color space
along the vortex, one encounters frustrations where one is forced to switch
to the negative 3-direction. At the frustrations, magnetic flux
corresponding to an Abelian monopole is supplied to or taken from the 
vortex, cf.~Fig.~\ref{vormon}.
Center vortex surfaces in Abelian gauges thus naturally contain Abelian
magnetic monopoles. In fact, they are necessary to generate a global
topological charge; on oriented vortex surfaces,
the local contributions to the topological charge globally add up to zero
\cite{cont}.

The real physical difference to the dual superconductor picture, cf.~section
\ref{monsec}, thus does not lie in the presence or absence of monopoles, but
in the form taken by the field strength emanating from those monopoles. In
the dual superconductor picture, a monopole is assumed to generate a radial
magnetic Coulomb field, whereas in the vortex picture, the field strength is
constricted into center vortex fluxes. Of course, in the latter picture,
the vortices are taken to be the primary degrees of freedom, and the
behavior of the monopoles on the vortices is seen as a consequence of
the primary vortex dynamics. In the dual superconductor, the monopoles
are assumed to be the primary degrees of freedom.

\subsubsection{Dynamics -- the random vortex world-surface model}
\label{randmod}
As discussed further above, it is important to test whether a particular
set of infrared collective gluonic degrees of freedom can indeed generate
the main features of the strong interaction vacuum on the basis of a weakly
coupled dynamics. To this end, a random vortex world-surface model has
been introduced and studied \cite{vormod,topol,csb,su3,bary}. In this
model, the vortex surfaces are composed of elementary squares on a
hypercubic lattice. The spacing of this lattice is a fixed physical
quantity interpreted as mimicking the effects of the finite vortex
thickness; parallel thick vortices cannot be distinguished anymore 
if they approach each other more closely than their radius, which turns
out to be roughly 0.4 fm (the scale is ultimately set by the zero-temperature
string tension). The lattice spacing at the same time provides the fixed
ultraviolet cutoff characteristic of any infrared effective model.
The random vortex surface ensemble is generated by Monte Carlo methods,
weighted by an action which controls vortex curvature at short distances.
On long distance scales, the vortex surfaces are thus random and uncorrelated.

On the basis of these assumptions, the random vortex world-surface model
reproduces both the confined and the deconfined phases of Yang-Mills
theory, for $SU(2)$ color \cite{vormod} as well as $SU(3)$ color \cite{su3},
in accordance with the heuristics presented further above. The
only dimensionless parameter of the model, namely the coupling strength
in the curvature action, is fixed by fitting the ratio of the
deconfinement temperature to the (square root of the) zero-temperature
string tension obtained in the corresponding Yang-Mills theory.
The model correctly predicts a second order deconfinement phase
transition for $SU(2)$ color and a very weakly first order transition
for $SU(3)$ color. Also the behavior of the spatial string tension in the
deconfined phase is described correctly. For $SU(3)$ color, baryonic static
quark configurations obey a Y area law \cite{bary}. Beyond the confinement
properties, also the $SU(2)$ topological susceptibility predicted by the model
agrees quantitatively with the one obtained in Yang-Mills theory
\cite{topol}, and chiral symmetry is broken spontaneously in the confined
phase, with a natural value for the (quenched) chiral condensate \cite{csb}.

\subsubsection{Vortex detection in lattice gauge configurations}
To study vortex physics in full lattice gauge theory, it is necessary to
devise methods to detect and isolate vortices in lattice gauge configurations.
Two main approaches have been studied in this respect. One approach relies
on gauges of the maximal center type \cite{giedt}, such as defined by the 
gauge condition (in the case of $SU(2)$ color)
$\max \sum_{i} \left| \mbox{tr } U_i \right|^{2} $,
where the $U_i $ are the link variables and the maximization is carried out
over all gauge transformations thereof. The gauge is thus chosen such as to
concentrate the link variables near the center elements of the gauge group.
In a second step, {\em center projection}, $U\rightarrow \mbox{sign tr } U$,
residual deviations away from those center elements are discarded, yielding
a $Z(2)$ gauge configuration, which is associated with vortex world-surfaces
on the dual lattice in canonical fashion (negative plaquettes in the $Z(2)$
configuration are pierced by vortices). Using this two step-procedure, one
aims to concentrate as much of the relevant physical information on the
collective center vortex degrees of freedom as possible.

After the introduction of maximal center gauges, it was noted that string
tensions measured in the projected vortex ensemble depend substantially
on the selection of Gribov copies when fixing the gauge \cite{bornyakov}. The
precise gauge fixing and projection procedure needs to be selected carefully
in order to ensure that the resulting projection vortex ensemble captures
the main infrared physics. While the situation in the case of $SU(3)$ color
is still unclear \cite{langsu3}, for $SU(2)$ color, a quite
successful procedure can be constructed, which in fact turns out to be a
hybrid \cite{hybrid} of maximal center gauge fixing and the second major
method of detecting vortices, namely the Laplacian center gauge.

The Laplacian center gauge \cite{lapcg} is free of Gribov copy selection
issues. It is defined via eigenvectors of the lattice adjoint Laplacian
operator. Two such eigenvectors are chosen, usually the ones with the
lowest eigenvalues; vortices are located on the two-dimensional submanifolds
of space-time where the eigenvectors become collinear in color space. In
addition, Abelian monopoles are located on the lines where the eigenvector
corresponding to the lowest eigenvalue vanishes. Thus, by construction,
the monopoles are located on the vortex surfaces in this approach.

Ultimately, it would be desirable to devise a gauge-invariant vortex
detection method. In principle, the desired information is contained
in the Wilson loops evaluated in a given gauge configuration; in
practice, extracting vortices from them is a difficult
pattern recognition problem, especially since the density of thick
vortices in the Yang-Mills vacuum is such that the vortices overlap
considerably; for a recent investigation which touches upon this issue,
cf.~\cite{chelu}.

\subsubsection{Results of lattice studies}
Based on the vortex detection procedures described above, a number of
vortex properties have been investigated. They include (the following are
results for $SU(2)$ color, except where otherwise indicated):
\begin{itemize}
\item
{\bf Center dominance of the string tension:} Extracting the vortex content
of lattice configurations and measuring Wilson loops in the resulting vortex
ensemble, one recovers the full Yang-Mills string tension 
\cite{hybrid,forpepe}. As mentioned above, when using the maximal
center gauge, one must be judicious in dealing with the Gribov problem.
\item
{\bf Vortex-limited Wilson loops:} Having identified the vortex content of
the Yang-Mills ensemble, one can select subensembles in which a large
region of a given lattice plane is not pierced by a center projection
vortex or in which this region is pierced by exactly one vortex.
Evaluating a Wilson loop in the aforementioned region using the full
Yang-Mills configurations contained in either subensemble, and taking the
ratio of the values obtained, one ends up with a ratio approaching $-1$ as
the Wilson loop becomes large \cite{hybrid}. This indicates that the thin
center projection vortex indeed represents the location of a thick center
flux in the full Yang-Mills configuration; when the Wilson loop becomes 
sufficiently large, it encompasses the full flux, yielding the appropriate
center phase. Effects of other fluctuations around this center flux are
largely canceled in the ratio.
\item
{\bf Vortex removal also removes all nonperturbative effects:} Center
projection yields a $Z(2)$ lattice configuration with nontrivial center
phases on selected links. Removing those center phases from the links
of the full Yang-Mills configuration defines the vortex-removed
configuration. In the vortex-removed ensemble, the string tension and
the chiral condensate vanish, and all configurations are in the
topological charge zero sector \cite{fordel}.
\item
{\bf Deconfinement transition as a percolation transition:}
At finite temperatures, the center projected vortex ensemble reproduces
the deconfinement phase transition, at which it displays a percolation
transition in accordance with the heuristics given further above
\cite{finitet,langce}.
\item
{\bf Roughening:} The center projection vortex density is suppressed within 
the chromoelectric flux tube between static sources, and can be used as
a measure of the width of that flux tube. Varying the distance between
the static sources, evidence of roughening is obtained \cite{rough1,rough2}.
\item
{\bf Maximal Abelian gauge monopoles on vortices:} While in the Laplacian
center gauge, monopoles are located on vortices by construction, one
also finds empirically that maximal Abelian gauge monopoles are located
on maximal center gauge vortices \cite{ambmon,mononv}.
Moreover, the action in the monopole region is not distributed isotropically,
but is collimated in the directions of the vortex.
\item
{\bf Topological susceptibility:} The center-projected vortex ensemble
generates a topological susceptibility compatible with the one obtained
in full Yang-Mills theory \cite{bertop}.
\item
{\bf Vortices and Higgs fields:} Extracting center projection vortices from
an $SU(2)$-Higgs ensemble, no vortex percolation is observed in the Higgs
(screening) phase \cite{berhiggs}.
\item
{\bf SU(3):} Generalizing to $SU(3)$ color, center dominance of the string
tension is again observed for the center vortices extracted via the
Laplacian center gauge \cite{forpepe}; significant deviations from center
dominance have been reported when gauges of the maximal center type are
used \cite{langsu3}, presumably due to the Gribov problems already
observed in the $SU(2)$ case.
\end{itemize}

\subsection{The dual superconductor}
\label{monsec}
Abelian magnetic monopoles are quantized sources and sinks of Abelian
magnetic flux. To be definite, one can select a $U(1)^{N-1} $ subgroup of
$SU(N)$ Yang-Mills theory by fixing the color direction of covariantly
transforming quantity of one's choosing; monopoles are sources and
sinks of flux with respect to that $U(1)^{N-1} $. The ensemble of
monopole world-lines in space-time can be cast in terms of a field theory.

\subsubsection{Confinement}
Motivated by the Ginzburg-Landau description of type-II superconductors,
in which {\em magnetic} flux is constricted into flux tubes, in the dual
superconductor picture one formulates a dual relativistic analogue, in
which {\em chromoelectric} flux is constricted, thus
generating confinement. The role of the (electrically charged) Cooper pairs
is taken over by Abelian magnetic monopoles. The corresponding
Ginzburg-Landau Lagrangean is the one of the dual Abelian Higgs model,
\[
{\cal L} \! = \! \frac{1}{4} G_{\mu \nu } G^{\mu \nu } +
\left| (\partial_{\mu } + igC_{\mu } ) \phi \right|^{2}
+\frac{\lambda }{4} (\phi \phi^{*} -v^2 )^2
\]
Here, $\phi $ describes the monopole field, and $C_{\mu } $ is a {\em dual}
$U(1)^{N-1} $ gauge field inducing a dual field strength $G_{\mu \nu } $.
This theory contains Nielsen-Olesen vortex solutions of {\em chromoelectric}
flux if the monopoles are condensed, i.e., if the monopole world-lines
percolate; thus, a confining potential between static quarks is induced.
When monopoles are not condensed, chromoelectric sources are
deconfined; both the confined and the deconfined phases can
therefore be described in a model based on Abelian magnetic monopoles.

However, due to the Abelian character of the description, adjoint $SU(N)$
sources, which are doubly charged with respect to the $U(1)^{N-1} $ gauge
group of the effective model, are also confined. One does not obtain the
correct n-ality dependence of string tensions in the dual superconductor
picture. As will be discussed in greater detail below, the crucial
truncation leading to this behavior lies in neglecting the off-diagonal
gluon degrees of freedom present in the underlying Yang-Mills theory.
This is often justified with the argument that these fields acquire a mass
and should therefore decouple. This appears to be fallacious, and to
restore the effect of the off-diagonal gluons in the dual superconductor,
one needs to add a coupling to charge-2 matter fields. These, indeed,
screen adjoint $SU(N)$ sources, restoring the correct n-ality dependence
of string tensions.

\subsubsection{Topological charge}
Topological charge implies the presence of magnetic monopoles, as is apparent
from the following observations:
\begin{itemize}
\item
Only non-oriented vortices (i.e., ones containing monopoles) carry global
topological charge.
\item
Instantons in Abelian gauges exhibit monopoles (and vortices).
\item
Calorons generically consist of monopole constituents.
\end{itemize}
However, conversely, monopole positions in general are not enough to
specify topological charge; one needs to specify the magnetic field
strength emanating from the monopoles (except in special gauges, such
as the Polyakov gauge). One particular such specification indeed generates
the correct topological susceptibility, namely collimating the magnetic
flux into vortices, as embodied in the random vortex world-surface model
\cite{topol}. Associating the monopoles with radial magnetic Coulomb fields
likewise yields topological charge; in an application to the monopoles
extracted from lattice Yang-Mills configurations, cf.~below, recovery of
$70 \% $ of the topological charge present in the full configurations was
reported \cite{sasaki}.

\subsubsection{Monopole detection in lattice configurations}
In principle, monopoles can be defined by finding the eigenvalues
in color space of a covariantly transforming quantity of one's choosing
(monopoles being located where two eigenvalues are degenerate); in practice,
the maximal Abelian gauge \cite{kronfeld}, defined by the gauge condition
$\max \sum_{i} \mbox{tr } \left( U_i \sigma^{3}
U_i^{\dagger } \sigma^{3} \right) $,
which renders the gauge fields as diagonal as possible, is used to 
determine the monopole content of a ($SU(2)$) lattice gauge configuration.
This gauge leaves a residual $U(1)$ gauge symmetry under which the diagonal
part of the gauge field transforms as a photon, whereas the off-diagonal
fields transform as charge-2 matter fields. In a second step, {\em Abelian
projection}, residual off-diagonal parts in the gauge field are discarded
and the diagonal fields are rescaled such as to preserve unitarity of
the link variables. Monopole positions can be identified by searching
for Dirac string fluxes leaving elementary lattice cubes. Beyond
Abelian projection, one can furthermore define {\em monopole projection},
in which only the information on the monopole positions is kept and
new configurations are constructed in which magnetic Coulomb fields are
associated with the monopoles \cite{shiba,stack}. This corresponds most
closely to the dual superconductor picture. It should be emphasized
that, in contrast to monopole projection, Abelian projection introduces
no a priori constraint on the form of the Abelian fields; the form of
these fields is still determined by the dynamics and they could,
e.g., be vortices. Indeed, being generated using the full Yang-Mills
dynamics, the gauge fields in the Abelian projected configurations
contain the effects of the off-diagonal gluon dynamics. By contrast, in
monopole projection, merely the monopole positions still contain the
effects of the full dynamics, whereas the field strength emanating from
the monopoles is Coulombic by construction.

Before continuing with the discussion of lattice results obtained in
practice using these projected configurations, it should be noted that
also further conceptual refinements have been proposed recently regarding the
consistent definition of electric and magnetic currents in the Abelian
projection framework \cite{haymaker}.

\subsubsection{Results of lattice studies}
A variety of issues have been investigated based on the Abelian and
monopole projection techniques discussed above. They include:
\begin{itemize}
\item
{\bf Abelian dominance:} Evaluating Wilson loops using Abelian projected
configurations yields $(92\pm 4) \% $ of the full fundamental string tension
\cite{yotsu,bali}. Adjoint sources are not confined.
\item
{\bf Monopole dominance:} Using monopole projected configurations to
compute Wilson loops, one recovers $87 \% $ of the full fundamental string
tension \cite{shiba,stack}. However, adjoint sources are confined
\cite{ambmon}.
\item
{\bf Percolation properties:} While the world-lines of maximal Abelian gauge
monopoles percolate in the confined phase, they cease to percolate in the
deconfined phase, both in the quenched case and in the case of two dynamical
quark flavors \cite{bornbar}.
\item
{\bf Flux tube properties:} Examining the structure of the monopole and
photon fields in the region of the chromoelectric flux tube between static
color sources, the parameters of the corresponding Nielsen-Olesen vortex
solution of the dual Abelian Higgs model can be identified. This has
been investigated both for the quenched case \cite{ano1,ano2,ano3,ano4}
and for the case of two dynamical quark flavors \cite{ano4}. No roughening
is reported.
\item
{\bf Baryonic configurations:} The potential of baryonic static quark
configurations evaluated using the Abelian projected ensemble obeys a
Y law \cite{bornbar}.
\item
{\bf Fundamental string breaking:} Fundamental string breaking at large
distances between static quark sources in the presence of two flavors of
dynamical quarks has been observed using Abelian projected configurations
\cite{stribrab}.
\item
{\bf Monopole action and entropy:} The effective action, world-line length
distribution and entropy have been extracted for maximal Abelian gauge
monopoles \cite{entrop}.
\item
{\bf Magnetic disorder parameter:} A magnetic disorder parameter has been
defined in order to study magnetic monopole condensation, in particular
its relation to the deconfining phase transition 
\cite{disord1,disord2,disord3,disord4,disord5,disord6}.
\item
{\bf Adjoint string breaking:} The dual superconductor incorrectly
confines adjoint color sources. This can be remedied by coupling the dual
Abelian Higgs model to charge-2 matter fields, which screen adjoint
sources \cite{ahm2}. Such a coupling is indeed natural, since the
underlying Yang-Mills theory contains off-diagonal gluon fields, which
transform as charge-2 matter fields in the maximal Abelian gauge. The
importance of this coupling has also been corroborated by the observation
that the charge-2 matter fields provide an essential part of the total
action in Abelian projected lattice Yang-Mills theory \cite{hashim}.
However, pursuing this train of thought further, the dynamics of the
charge-2 matter fields have the additional effect that monopoles are
arranged into chain-like structures \cite{schiller}, which can be
identified as vortices \cite{greensrev}. Thus, restoring the correct
n-ality dependence of string tensions in the dual superconductor model
by introducing charge-2 matter fields ultimately guides one towards
adopting the vortex picture.
\item
{\bf Monopole world-line correlations:} The world-lines of maximal Abelian
gauge monopoles are not random walks on infrared scales, but exhibit
long-range correlations characteristic of the monopoles being associated with
two-dimensional surfaces in space-time \cite{boyko}. This observation is
consistent with the fact that maximal Abelian gauge monopoles are located
on center vortex world-surfaces, the dynamics of which determine monopole
behavior.
\end{itemize}

\subsection{Topological charge lumps}
\subsubsection{Merons}
A ($SU(2)$ color) meron located at the space-time origin can be described
by the gauge field
\[
a_{\mu } (x) = \frac{\eta_{a\mu \nu } x_{\nu } }{x^2 + \rho^{2} }
\frac{\sigma^{a} }{2}
\]
with the 't~Hooft symbol $\eta_{a\mu \nu } $. For width parameter $\rho =0$,
this is a (albeit singular) solution of the classical Yang-Mills equations
of motion. Merons carry topological charge $\pm 1/2$; since their field
strength behaves as $1/x^2 $ at large distances $x$, the action of a
single meron is logarithmically divergent. However, one can construct a
meron gas with finite action density if the meron color orientations are
suitably correlated \cite{lenz}. Using the ansatz
\[
A_{\mu } (x) = \sum_{i=1}^{M} h_i a_{\mu } (x-z_i ) h_{i}^{-1}
\]
for a configuration of $M$ merons at positions $z_i $, where the $h_i $
are color rotation matrices, one can define a meron ensemble by summing over
all $z_i $ and $h_i $ weighted with the
Yang-Mills action. In practice, the ensemble is created via Monte Carlo
methods; note that the update is non-local, since any given meron interacts
with all other merons. This model induces nontrivial color correlations
between the merons, which result in a confining linear potential between
static quarks. Using the string tension to fix the scale, one simultaneously
finds a topological susceptibility which agrees quantitatively with the
one obtained in $SU(2)$ Yang-Mills theory. It would be interesting to
further investigate the correlations between the merons, in particular with
a view to understanding whether higher-dimensional collective excitations
are induced by the dynamics of the model.

\subsubsection{Instantons}
Instantons are regular solutions of the classical Yang-Mills equations of
motion which are localized in space-time and carry topological
charge $\pm 1$. They are the basis for the highly successful instanton
liquid model of the strong interaction vacuum, which describes the
$U_A (1)$ anomaly, the spontaneous breaking of chiral symmetry, and a
wealth of hadron phenomenology associated with those effects, cf.~the
review \cite{shuryak}. However, there is no confinement in the instanton
liquid model.

A detailed discussion of methods used for detecting instantons in lattice
gauge configurations constitutes a separate topic which lies beyond the
scope of this review. Tests of the instanton picture within the framework
of lattice gauge theory have revealed the following results \cite{negele}:
\begin{itemize}
\item
The procedure of cooling lattice configurations can be used as a filter to
retain only classical solutions, i.e., instantons. Using cooled
configurations to evaluate light hadron correlators, one obtains
essentially the same light hadron phenomenology as when using the
full lattice configurations. On the other hand, confinement disappears.
\item
There are correlations between the positions of low virtuality quark modes
in the full configurations and the positions of instantons in the cooled
configurations.
\item
Low virtuality quark modes dominate light hadron physics. Truncating the
spectrum of the Dirac operator in the lattice gauge ensemble to include
only these modes approximates hadronic correlators well.
\end{itemize}
These findings suggest viewing the low virtuality quark modes dominating
light hadron phenomenology as being generated by instantons in the strong
interaction vacuum. Indeed, an isolated instanton gives rise to a zero mode
of the Dirac operator, and an ensemble of not completely isolated instantons
therefore generates a band of near-zero modes (thus inducing the spontaneous
breaking of chiral symmetry via the Casher-Banks relation). On the other
hand, it is just as consistent with the above observations that a different
class of topological charge lumps induces the low virtuality quark modes,
and that replacing these by modes generated by instantons represents only
a mild (and very useful) idealization. Note that, even if such a different
class of topological charge lumps is present in the full lattice
configurations, it will be reduced to instantons under cooling, since the
latter are minima of the Yang-Mills action. The fact that the instantons in
the cooled lattice configurations faithfully reproduce only partial aspects
of the strong interaction vacuum, namely the correct ensemble of
low virtuality quark modes, but not confinement, points toward the
relevant degrees of freedom carrying topological charge in that vacuum
being different from instantons.

Motivated by the strongly correlated confining meron model described in
the previous section, a completely analogous strongly correlated instanton
model has been investigated very recently \cite{neghere}. This model indeed
also produces confinement of fundamental color sources, and, moreover, does
not confine adjoint color sources \cite{negpriv}. Again, it would be
interesting to further investigate the correlations between the instantons,
especially whether higher-dimensional collective excitations are induced
by the dynamics of the model. Also modifications of instanton liquid
phenomenology by the color correlations present in this model need to be
investigated.

\subsubsection{Calorons}
Calorons are solutions of the classical Yang-Mills equations of motion
on space-times with compact directions, such as the ones used to describe
finite temperatures; they carry quanta of topological charge. Calorons
display striking space-time properties, generically consisting of
monopole constituents, where it should be emphasized that these
monopoles are defined in a fully gauge-invariant fashion \cite{vbaal1,vbaal2}.
No gauge fixing procedure is necessary to identify the constituents in
the caloron solutions. It seems tempting to speculate that these monopoles
may be associated with confinement in a caloron ensemble.

Calorons and their monopole substructure have recently been detected in
lattice gauge configurations via cooling methods \cite{ilgenf} as well
as via associated quark zero modes \cite{gatt1,gatt2,gatt3}. The
observations in these studies suggest that the topological charge
distribution in the QCD vacuum is more fragmented and structured than
in an ensemble of uncorrelated instanton-like lumps. This is corroborated
by investigations which indicate the presence of a long-range low-dimensional
topological charge structure in the QCD vacuum \cite{horvath}.

\section{SYNOPSIS}
Of the collective infrared gluonic degrees of freedom considered, center
vortices are the only ones for which a weakly coupled model has been
been formulated which reproduces the main nonperturbative features of
the strong interaction vacuum. This supports the notion that center
vortices are the actual relevant degrees of freedom in the infrared sector.
Going from the vortex picture to the dual superconductor picture and to
weakly coupled models based on topological charge lumps, an increasing
level of truncation of the degrees of freedom and the associated
nonperturbative physics is introduced. Conversely, if one wishes to
nevertheless describe the full spectrum of nonperturbative effects using
Abelian monopoles or topological charge lumps, one is forced to formulate
models with progressively stronger correlations to recover all the relevant
physics. This is a signature that those degrees of freedom are really
combined into other collective degrees of freedom which more accurately
characterize the strong interaction vacuum. Nevertheless, if one wishes
to focus on a particular type of nonperturbative effect, the more
strongly truncated models can be very useful; this is best evidenced by
the wealth of phenomenological results obtained in the instanton liquid
model, unparalleled by any of the other approaches, despite the absence of
confinement in this model.

In detail, recent work has led to a convergence of the vortex and the dual
superconductor pictures. Both contain Abelian monopoles; however,
in the vortex picture, they are a secondary consequence of the
nonorientability of vortex world-surfaces and their behavior is determined
by the primary vortex dynamics. As a case in point, monopoles inherit the
percolation properties of the vortex surfaces on which they are located;
the deconfinement transition can be equally detected as a percolation
transition in the monopole world-lines or the vortex world-surfaces. Going
from the vortex picture to the dual superconductor picture, one introduces
two truncations. On the one hand, the field strength sourced at the monopoles
is assumed to take a radial Coulomb form instead of being constricted into
vortex fluxes; on the other hand, monopole world-lines are assumed to behave
as random walks in space-time instead of displaying correlations
characteristic for them being located on two-dimensional random surfaces.
The former truncation in particular leads to the loss of the correct n-ality
dependence of string tensions; not only fundamental color sources, but also
adjoint ones are confined in the dual superconductor. Moreover, lattice
studies have indeed shown that monopole world-lines do not behave as random
walks, but do display correlations characteristic of them being located on
vortex surfaces \cite{boyko}. To remedy these truncations, one either has to
revert to the vortex picture or, which is presumably equivalent, resort to
more complicated models of Abelian monopoles coupled to charge-2 matter
fields (which are descendants of the off-diagonal gluons in Yang-Mills
theory). This coupling, which is discarded in the standard dual
superconductor scenario, is evidently instrumental in constricting
chromomagnetic flux into vortices, thus inducing the correct n-ality
dependence of string tensions \cite{schiller,greensrev}.

The connection between the vortex picture and models based on topological
charge lumps is not yet understood in quite as much detail. Center vortices
contain topological charge lumps induced by vortex writhe and vortex
self-intersections, cf.~Fig.~\ref{topex}. It seems plausible that the physics
controlled by this topological density can be well described by models
based directly on topological charge lumps; indeed, the instanton liquid
model is very successful in describing the effects of spontaneous chiral
symmetry breaking and the axial $U_A (1)$ anomaly. Nevertheless, by
formulating weakly coupled models of this type, which, e.g., do not include
any correlations characteristic of the topological charge being located on,
and induced by, vortices, one is evidently truncating relevant physics;
confinement is lost. Accordingly, evidence from lattice studies is mounting
that the topological charge distribution in the QCD vacuum is indeed more
fragmented and structured than suggested by a picture of uncorrelated
instanton-like lumps \cite{gatt3,horvath}; topological charge must
be organized into higher-dimensional long-range collective degrees of
freedom to recover the full physics. This is also indicated by the recent
construction of strongly correlated meron and instanton models which do
display confinement \cite{lenz,neghere}. However, the precise nature of
these correlations and the collective degrees of freedom induced by them
remains to be studied in more detail; the aforementioned new meron and
instanton models may turn out to be very valuable laboratories in
this respect. They indicate that, in particular, the color orientations
of topological charge lumps must be properly aligned.

Finally, the intriguing cross-connection between calorons and monopoles
should be noted; calorons are not unstructured lumps of topological charge,
but generically consist of monopole constituents, where it should be
emphasized that these monopoles are defined in a gauge-invariant manner,
as opposed to the maximal Abelian gauge monopoles the dual superconductor
scenario usually refers to. Nevertheless, this connection suggests that
confinement in the caloron picture may possibly be generated through
these monopole constituents. It would also be interesting to
understand in detail whether, and how, vortex flux may enter the caloron
picture.

\section*{Acknowledgments}
\vspace{-0.23cm}
The author is grateful to the Local Organizing Committee and the
International Advisory Committee for the opportunity to present this review,
and to the many colleagues who have furnished their insights on diverse
topics touched upon here, in particular J.~Negele and P.~van~Baal. This 
work was supported by the U.S.~DOE under grant number DE-FG03-95ER40965.

\end{document}